\documentclass[twoside]{dis08}
\usepackage[latin1]{inputenc}
\usepackage[dvips]{graphicx,epsfig,color}
\usepackage{wrapfig,rotating}
\usepackage{amssymb,amsmath,array}

\def\gtap{\raisebox{-.4ex}{\rlap{$\sim$}} \raisebox{.4ex}{$>$}}

\newcommand {\pom} {I\!\!P}

\begin{document}
\title{Observation of single-diffractive $W$ production with CMS: a feasibility study}

\author{A. Vilela Pereira, on behalf of the CMS Collaboration
%
%
\vspace{.3cm}\\
%
Universit\` a degli Studi di Torino \& INFN Torino\\
Via Pietro Giuria 1, 10125 Torino, Italy
%
}

\maketitle

\begin{abstract}
We present a study of single-diffractive $W$-boson production in $pp$
collisions at $\sqrt{s}=14$~TeV, $pp \to Xp$ with $X$ including a $W$
boson, with the CMS detector.  We discuss the feasibility of observing
this process with an integrated effective luminosity for
single interactions of 100~pb$^{-1}$.
\end{abstract}

\section{Introduction}

\begin{wrapfigure}{r}{0.5\columnwidth}
\centerline{\includegraphics[width=0.45\columnwidth]{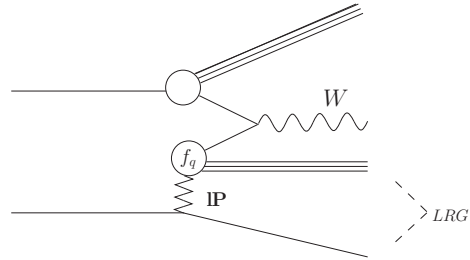}}
\caption{Sketch of the single-diffractive
reaction $pp \to Xp$ in which  $X$ includes a $W$ boson. The symbol
$\pom$ indicates the exchange with the vacuum quantum numbers (Pomeron).
The large rapidity gap (LRG) is also shown.}\label{Fig:diagram}
\end{wrapfigure}

In the present paper, the single-diffractive (SD) reaction $pp \to Xp$ is
studied, in which $X$ includes a $W$ boson (Fig.~\ref{Fig:diagram}). The
$W \to \mu \nu$ decay mode is considered. This reaction is sensitive to
the diffractive structure function of the proton and to the rapidity gap
 survival probability~\cite{Bj}. This process has been
studied at the Tevatron, where the ratio of the yields for SD and
inclusive $W$ production has been measured to be approximately
1\%~\cite{Abe:1997jp,Abazov:2003ti}.

The aim of this analysis is to demonstrate the feasibility of observing SD
$W$ production at CMS given an integrated effective luminosity for single
interactions of 100~pb$^{-1}$; this effective luminosity will be lower
than the integrated delivered luminosity and will depend on the machine conditions.

The CMS apparatus is described in detail elsewhere~\cite{PTDR1}. Two
experimental scenarios are considered here. In the first, no forward
detectors beyond the CMS forward calorimeter HF, which covers the region
in pseudo-rapidity of $3<|\eta|<5$, are assumed. In the second,
additional coverage at $-6.6<\eta<-5.2$ is assumed by means of the CASTOR
calorimeter.

\section{Monte Carlo Simulation}

%
%
Single diffractive $W$ production was simulated by using the {\sc pomwig}
generator~\cite{Cox:2000jt}, version v2.0 beta. For the diffractive PDFs and the Pomeron flux, the result of
the NLO H1 2006 fit B~\cite{Aktas:2006hy} was used. A rapidity gap survival probability of 0.05, as predicted
in~\cite{Khoze:2006gg}, is assumed.  The cross section, for the $W \to
\mu\nu$ mode, is about 70~pb, leading to $\simeq 7000$ events per
100~pb$^{-1}$. For non-diffractive $W$ production, the {\sc pythia} generator was
used~\cite{Sjostrand:2006za}. The cross section, for the $W \to \mu\nu$
mode, is about 22 nb (NLO).  With the given numbers for the cross
sections, the ratio of diffractive to inclusive yields is 0.3\%.

\section{Event Selection and Observation of SD $W$ Production}
\label{selection-observation}

\subsection{$W \to \mu \nu$ selection}
\label{inclusive-selection}

The selection of the events with a candidate $W$ decaying to $\mu \nu$ is
the same as that used in~\cite{inclusive}.  Events with a candidate muon
in the pseudo-rapidity range $|\eta|>2.0$ and transverse momentum
$p_T<25$~GeV were rejected, as were events with at least two muons with
$p_T>20$~GeV. Muon isolation was imposed by requiring $\sum{p_T}<3$~GeV in
a cone with $\Delta R<0.3$.  The transverse mass was required to be
$M_T>50$~GeV. The contribution from top events containing muons was
reduced by rejecting events with more than 3 jets with $E_T>40$~GeV
(selected with a cone algorithm with radius of 0.5) and events with
acoplanarity ($\zeta=\pi- \Delta{\phi}$) between the muon and the
direction associated to $E_T^{\rm miss}$ greater than 1 rad. Approximately
2,400 SD $W$ events and 600,000 non-diffractive $W$ events per
100~pb$^{-1}$ are expected to pass these cuts.

\subsection{Diffractive selection}

\begin{wrapfigure}{r}{0.5\columnwidth}
\centerline{\includegraphics[width=0.45\columnwidth]{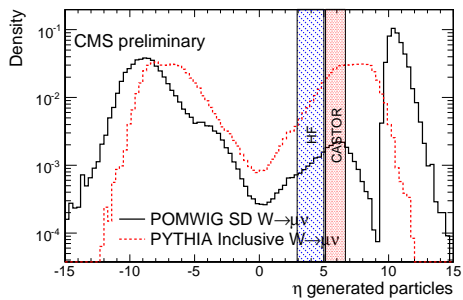}}
\caption{
Generated energy-weighted $\eta$ distribution for stable particles
in diffractive ({\sc pomwig}, continuous line) and
non-diffractive ({\sc pythia}, dashed line) events.
 The diffractive events were
generated with the gap side in the positive $\eta$ hemisphere.
}
\label{Fig:etagen}
\end{wrapfigure}

Figure~\ref{Fig:etagen} shows the generated
energy-weighted $\eta$ distribution for stable particles (excluding
neutrinos) in diffractive and non-diffractive events, including the
scattered proton; all events were generated with the scattered proton at
positive rapidities (the peak at $\eta \gtap 10$). Diffractive events
have, on average, lower multiplicity both in the central region and in the
hemisphere that contains the scattered proton, the so-called ``gap side",
than non-diffractive events.

The gap side was selected as that with lower energy sum in the HF.
A cut was then placed on the multiplicity of tracks with $p_T>900$~MeV and $|\eta|<2$.
 For the events passing this cut, multiplicity distributions in
the HF and CASTOR calorimeters in the gap side were studied, from which a diffractive
sample can be extracted.

\subsection{Evidence for SD $W$ Production}
\label{evidence}

\subsubsection{HF multiplicity}

\begin{wrapfigure}{l}{0.6\columnwidth}
\centerline{\includegraphics[width=0.55\columnwidth]{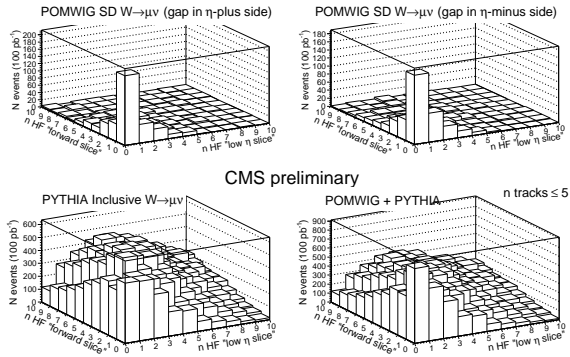}}
\caption{
Low-$\eta$ (``central slice") vs high-$\eta$ (``forward
slice") HF tower multiplicity distributions for events with
track multiplicity in the central tracker $N_{\rm track} \le 5$.
}
\label{Fig:HF-HF-NT5}
\end{wrapfigure}

Figure~\ref{Fig:HF-HF-NT5} shows the HF tower multiplicity for the low-$\eta$
(``central slice", $2.9 <\eta<4.0$) and high-$\eta$ HF (``forward slice",
$4.0<\eta<5.2$) regions for events with central track multiplicity
$N_{\rm track} \le 5$. The top
left and top right plots show the distributions expected for the
diffractive $W$ events with generated gap in the positive and negative $Z$
direction, respectively; they exhibit a clear peak at zero multiplicity.
Conversely, the non-diffractive $W$ events have on average higher
multiplicities, as shown in the bottom left plot. Finally, the bottom
right plot shows the sum of the {\sc pomwig} and {\sc pythia}
distributions -- this is the type of distribution expected from the data.
The diffractive signal at low multiplicities is visible.

\subsubsection{HF and CASTOR Multiplicity Distributions for the Gap Side}

The HF tower multiplicity vs CASTOR $\phi$ sector multiplicity was studied
for the gap side.  Since CASTOR will be installed at first on the negative
side of the interaction point, only events with the gap on that side (as
determined with the procedure discussed above)  were considered. The CMS
software chain available for this study did not include
simulation/reconstruction code for CASTOR; therefore, the multiplicity of
generated hadrons with energy above a 10~GeV threshold in each of the
CASTOR azimuthal sectors was used.

\begin{wrapfigure}{r}{0.6\columnwidth}
\centerline{\includegraphics[width=0.55\columnwidth]{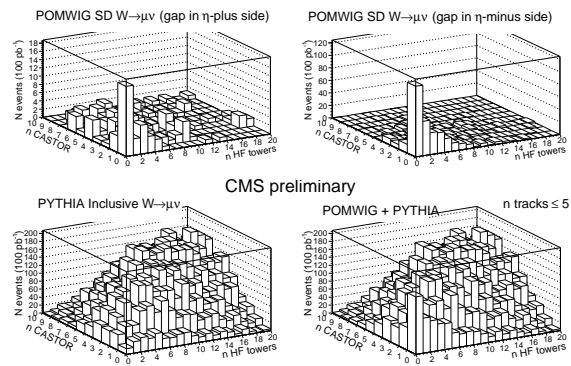}}
\caption{HF tower multiplicity
vs CASTOR sector multiplicity distribution for events with
track multiplicity in the central tracker $N_{\rm track} \le 5$.
}
\label{Fig:HF-CASTOR-NT5}
\end{wrapfigure}

Figure~\ref{Fig:HF-CASTOR-NT5} shows plots analogous to those of
Fig.~\ref{Fig:HF-HF-NT5} for the combination of HF and CASTOR. The top plots show the {\sc pomwig} distributions; the few
events in the top left plot are those for which the gap-side determination
was incorrect.  The signal to background ratio improves greatly with
respect to the HF only case.

\subsection{Observation of SD $W$ Production and Extraction of a 
Diffractive Sample}

The two-dimensional multiplicity plots of Figs.~\ref{Fig:HF-HF-NT5}
and~\ref{Fig:HF-CASTOR-NT5} provide
evidence of SD $W$ production. A simple way to isolate a sample of
diffractive events from these plots is to use the zero-multiplicity bins,
where the diffractive events cluster and the non-diffractive background is
small.

The HF plus CASTOR condition yields the best signal to background ratio.
 When an integrated effective luminosity for single
interactions of 100~pb$^{-1}$ becomes available, SD $W \to \mu \nu$
production can then be observed with $\mathcal{O}(100)$ signal events.
Backgrounds other than non-diffractive $W$ production are
discussed in the next section and appear to be under control.

\subsection{Backgrounds}
\label{sec-bkd}

The background from QCD events has been quantified by repeating the analysis
 on a {\sc pythia} sample with lepton-enriched QCD events. The
effect is to increase the number of events in the low-multiplicity region
by less than 1\% of the SD yield. Non-diffractive $W$ events, misidentified as diffractive, have been discussed in the previous sections, where we showed they can be kept under control.

SD $W$ production with proton-dissociation, $pp \to XN$, where $X$
contains a $W$ boson and $N$ is a low-mass state into which the proton has
diffractively dissociated, is an irreducible background when $N$ escapes undetected in the forward region. A study of proton-dissociation has been carried out in~\cite{jonathan}, where it has been shown that about 50\% of the proton-dissociative background can be rejected by vetoing events with activity in the CMS Zero Degree Calorimeter (ZDC), which provides coverage for neutral particles for $|\eta| > 8.1$. The net effect is to enhance the diffractive signal in the zero multiplicity bin of Fig.~\ref{Fig:HF-CASTOR-NT5} by about 30\%.

\section{Summary and Outlook}

A procedure has been discussed to arrive at the observation of single
diffractive $W \to \mu \nu$ production with an integrated effective
luminosity for single interactions of 100~pb$^{-1}$. The procedure is
based on the detection of large rapidity gaps in the final state of the
event using HF and CASTOR, complemented by the multiplicity information 
from the central tracker.

Assuming a rapidity gap survival probability of 0.05,
$\mathcal{O}(100)$ reconstructed signal events are expected with a
high signal-to-background ratio if the CASTOR calorimeter is
available. If CASTOR is not available, the HF information alone may
be sufficient. Further improvements are possible if the detectors of
the TOTEM experiment~\cite{totem} can be used, notably the T2
tracker and the roman pot detectors.



\begin{footnotesize}

\end{footnotesize}


\end{document}